\documentclass[conference]{IEEEtran}

\usepackage{times}
\usepackage{amsmath}
\usepackage{amssymb}
\usepackage{algorithm}
\usepackage{algorithmic}
\usepackage{graphicx}
\usepackage{subfigure}
\usepackage{verbatim}
\usepackage[usenames]{color}
\usepackage{cite}
\usepackage{url}
\usepackage{soul}
\usepackage{todonotes}

\def\a{\mathbf{a}}

\def\x{\mathbf{x}}

\newtheorem{theorem}{Theorem}

\newtheorem{remark}[theorem]{Remark}
\begin{document}

\title{Non-line-of-sight Node Localization based on Semi-Definite Programming in Wireless Sensor Networks }

\author{
Hongyang Chen$^{1}$, Kenneth W. K. Lui$^{2}$, Zizhuo Wang$^{3}$, H. C. So$^{2}$, and H. Vincent Poor$^{4}$\\
         $^{1}$Institute of Industrial Science, The University of Tokyo, Tokyo, Japan\\
         $^{2}$Department of Electronic Engineering, City University of Hong Kong, Hong Kong\\
         $^{3}$Department of Management Science and Engineering, Stanford University, Standford, CA, USA\\
         $^{4}$Department of Electrical Engineering, Princeton University, Princeton, NJ, USA\\}

\maketitle

\begin{abstract}
An unknown-position sensor can be localized if there are three or more anchors making time-of-arrival (TOA)
measurements of a signal from it. However, the location errors can be very
large due to the fact that some of the measurements are from
non-line-of-sight (NLOS) paths. In this paper, we propose a semi-definite programming (SDP)
based node localization algorithm in NLOS environment
for ultra-wideband (UWB) wireless sensor networks. The positions of
sensors can be estimated using the distance
estimates from location-aware anchors as well as other sensors. However, in the absence of
LOS paths, e.g., in indoor
networks, the NLOS range estimates can be significantly biased. As a result, the NLOS
error can remarkably decrease the location accuracy.
 And it is not easy to efficiently distinguish LOS from NLOS measurements. In this paper, an algorithm is proposed that achieves high location accuracy without the need of identifying NLOS and LOS measurement.
\end{abstract}

\section{Introduction}
\label{sec:intro}

Localization algorithms for wireless sensor networks (WSNs) have been designed to find sensor location information, which is a major requirement in many applications of WSNs. Examples of such applications include animal tracking, mapping and location-aided routing.

Generally speaking, based on the type of information provided for localization, protocols can be divided into two categories: (i) range-based and (ii) range-free protocols \cite{apit}. Due to the coarse location accuracy, solutions of range-based localization are often more preferable and accurate than those of range-free schemes.
Range estimates from anchors can be obtained
using received signal strength (RSS), angle-of-arrival (AOA) or time-of-arrival (TOA)
observations of transmitted calibration signals \cite{mdtoa}. Impulse-based ultra-wideband
(UWB) is a promising technology where precise ranging can be embedded into data communication, due to its robustness in dense multipath environments and its ability to provide
accurate position estimation with low-data-rate communication.
In this paper, we focus on the investigation of range-based localization algorithms for UWB WSNs. One of the main challenges for accurate node localization in range-based localization algorithms is non-line-of-sight (NLOS) propagation which is caused by the obstacles in the direct paths of beacon signals. NLOS will
result in unreliable localization and significantly decrease the location accuracy if its effects are not taken into account. This often occurs in an urban or indoor environment. Some localization algorithms that cope with  the existence of NLOS range measurement have been proposed \cite{chanNLOS} \cite{congNLOS} \cite{wangNLOS}, mostly in cellular networks. Roughly speaking, there are two categories of approaches to deal with the localization problem in the presence of NLOS propagation. The first approach identifies LOS and NLOS information and discards the NLOS range information for position estimation. The second approach uses all NLOS and LOS measurements and provides weighting or scaling to reduce the adverse impact of NLOS range errors on the accuracy of location estimates, they also assume that the NLOS range estimates have been identified.

The number of anchors is typically limited by practical considerations. It might be a waste of resources to discard NLOS range measurements. To make best use of all range measurements, a computationally efficient semi-definite programming (SDP) approach that effectively incorporates both LOS and NLOS range information into the estimate of a sensor's location is proposed in this paper. We focus on the problem of NLOS mitigation, but do not require to accurately distinguish between LOS and NLOS range estimates. Given a mixture of LOS and NLOS range measurements, our method is applicable in both cases without discarding any range information. This method is the only SDP based approach to reduce the impact of NLOS on node localization in WSNs.
The main advantages of this approach are as follows.

1) The statistics of the NLOS bias errors are not assumed to be known \emph{a priori}.

2) No range information is discarded.

3) NLOS range estimates are not required to be readily distinguished from LOS range estimates through channel identification.


In our proposed approach, we assume the following features of UWB TOA-based range estimation: the range bias errors in NLOS conditions are always positive and significantly larger in magnitude than the range-measurement errors in LOS conditions.
In the next section, we show that the problem of node localization, given range information, can be cast into a nonlinear programming. We then use SDP relaxation techniques and add an additional measurement error to the actual range measurements, resulting in a method that suits for both LOS and NLOS range estimates to estimate a sensor's location.

The rest of the paper is organized as follows. Section II derives the SDP based localization algorithm in NLOS environment. In Section III, an extension model was proposed to deal with case when the anchor positions are also uncertain. In section IV, simulation results are reported. Section V draws the conclusion.

\section{NLOS Localization using SDP}
In this section, an SDP based node localization approach is proposed. We first introduce the technical preliminaries of this algorithm in subsection A and then formulate the localization problem as a nonlinear optimization problem in subsection B. An extension for our SDP algorithm to the case considering the anchor uncertainties is given in Section III.

\subsection{Background}
The basic setting of this paper is as follows: There are $n$ distinct sensors in $R^2$ whose positions are to be determined and $m$ anchors whose positions are known \emph{a priori}. We use $x_i\in R^2, i=1,2,...,n$ to denote the sensors and $x_j\in R^2, j=n+1,n+2,...,n+m$ to denote the anchors. We use $r_{i,j}$ to denote the actual distance between anchor and sensor or sensor and sensor, i.e.,
\begin{eqnarray}
{r_{i,j}} = \left\| {\x_i - \x_j } \right\|,\forall i=1,2,...,n, j=1,2,...,m+n.
\end{eqnarray}

In practice, we get measurement information for a subset of pairs of nodes, which we denote by $E$. We use $E_1$ to denote the measurement information between sensors and anchors and $E_2$ to denote the measurement information between sensors and sensors. By definition, $E=E_1\bigcup E_2$. Notice that this measurement could be either LOS or NLOS, and since we do not distinguish these two measurements, we do not need to separate $E$ by the type of measurement.

\begin{figure}[t]
\centering
\includegraphics[width=3.0in]{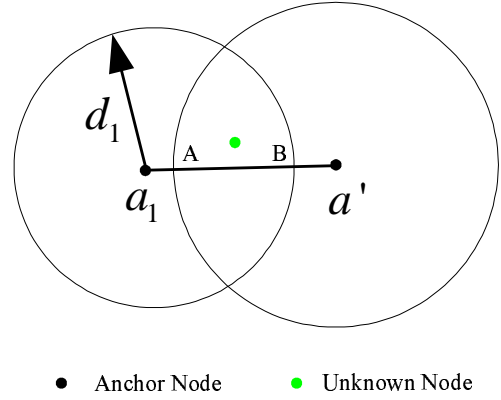}
\caption{An instance with two anchors} \label{fig:RSSI measurements}
\end{figure}

In this paper, we assume that the LOS range measurement is
\begin{eqnarray}
{d_{i,j}} = {r_{i,j}} + {n_{i,j}},
\end{eqnarray}
where ${n_{i,j}} \sim N(0,\sigma _{i,j}^2)$ is the measurement error which follows a zero-mean Gaussian distribution with standard deviation $\sigma_{i,j}$.

Similarly, the NLOS range measurement is assumed to be
\begin{eqnarray}\label{eqn:noise_model}
{D_{i,j}} = {r_{i,j}} + {n_{i,j}} + {\delta_{i,j}},
\end{eqnarray}
where $\delta _{i,j}$ is the error of NLOS measurement.

The idea of our approach is to get an upper bound as well as a lower bound for the true distance of each pair of nodes (could be either anchor and sensor or sensor and sensor) based on the measurement we observe, without distinguishing whether it comes from LOS or NLOS measurement. These bounds will form a feasible region for possible locations of each sensor and we then choose one ``center'' point from this region as our estimation.

First we show how we obtain the upper bound for the distance of certain pair of nodes.
Since in the NLOS case, the measured distance is larger than the actual distance, the measurement itself is an upper bound. For the LOS case, we have the upper bound as:
$$r_{i,j}\le d_{i,j}+n_{i,j}^U$$
where $n_{i,j}^U$ is an upper bound on the measurement error, which could be calculated in advance
based on experimental measurements. Therefore, we have a uniform upper bound for the distance between each pair of nodes as follows:
\begin{equation}
r_{i,j}=\left\|\x_i-\x_j\right\|\le d_{i,j}+n_{i,j}^U\mbox{   }\forall (i,j)\in E.
\end{equation}

Next we derive the lower bounds. Here we use the same idea as it is in \cite{caffery}. We first consider the distances between sensors and anchors. We consider those sensors that have more than two anchors in range. For each of those sensors, we draw a circle for each anchor in range centered at the anchor position and the radii is the upper bound computed by the above method. Obviously, these circles will have a common intersection part which contains the true location of the sensor. Now we look at each pair of the circles. As shown in Fig. 1, a lower bound of the distance between the sensor and anchor 1 is $\left\|d_1-AB\right\|$, where $AB$ is the intersection of the line connecting the centers of the two circles and the common area of the two circles. Then we take the maximum of this bound over all the anchors in range, and get a final lower bound as follows:

$$r_{i,j}\ge l_{i,j}\doteq\max_{k,(i,k)\in E_1}\left\|d_j-AB_k\right\|$$
for any pair of sensor and anchor in range. In the above formulation $d_j$ is the radii for circle $j$ and $AB_k$ is the line segment based on the intersection of circles $j$ and $k$. Therefore, for each pair of sensor and anchor in range, we get a lower bound for their distances.

However, for the distances between sensors and sensors, we cannot apply the same technique because the position of the sensors are not known. Thus for those sensors, the lower bound is set to zero.

Therefore, we get the following upper and lower bounds:

For each in-range sensors and anchors, we have
$$ l_{i,j}\le r_{i,j}\le d_{i,j}+n_{i,j}^U$$
and for each in-range sensors and sensors, we have
$$ 0\le r_{i,j}\le d_{i,j}+n_{i,j}^U.$$

For the later convenience, we uniformly write the upper and lower bound for the distance between node $i$ and $j$ by $u_{i,j}$ and $l_{i,j}$, respectively.

\begin{remark}
 In some circumstances, we do not have the communication between sensors and sensors. In that case, we simply remove the constraint between sensors, only keeping those between sensors and anchors.
\end{remark}
\begin{remark}
 This approach can also be applied to the cases when we have prior information on which measurement is from LOS path and which is from NLOS path. If we know \emph{a priori} that a certain measurement is from NLOS path, then we can compute the upper bound by simply using the measurement, or if we know the error is in a certain distribution, then we can again adjust the upper and lower bound accordingly. The same thing applies when we know a certain measurement is from LOS path.
\end{remark}

\subsection{Localization Algorithm using Semi-definite Programming for both NLOS and LOS Environments}
In this section, we present a convex optimization algorithm for node localization based on the bounds we obtained in the previous subsection.

As shown in Fig. 2, for the single constraint case ($s < \left\| {\x - \a} \right\| \leqslant R$), it is easy to see that one heuristic position estimate lies on the circle with center $\a$ and radius $\frac{{R + s}}{2}$. (By way of example, the square indicates the possible position for an efficient position estimate in Fig. 2.)

\begin{figure}[t]
\centering
\includegraphics[width=3.0in]{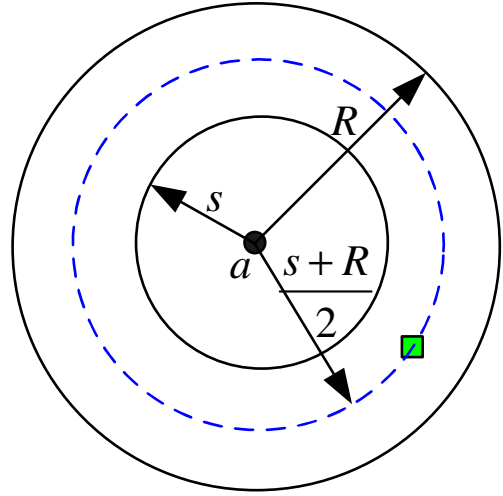}
\caption{The single constraint} \label{fig:RSSI measurements}
\end{figure}

This can be determined by minimizing the following expression:
\begin{eqnarray}\label{eqn:7}
{({\left\| {\x - \a} \right\|} - s)^2} + {({\left\| {\x - \a} \right\|} - R)^2}.
\end{eqnarray}
On expanding, (\ref{eqn:7}) becomes
\begin{align}\label{eqn:7_expand}
\begin{split}
    &(\| \x - \a \| - s)^2 + (\|\x - \a \| - R)^2\\
    &=2\|\x - \a\|^2 - 2(s+R)\|\x - \a\| + s^2 +R^2
\end{split}
\end{align}
where $s^2$ and $R^2$ are constants defined in the previous subsection.

Therefore, the optimization problem for locating the sensors can be formulated as:

\begin{equation}
\begin{array}{ll}\label{eqn:minproblem}
\mbox{min}{x} & \sum_{i<j:(i,j)\in E}[\left\|\x_i-\x_j\right\|^2-2(l_{i,j}+u_{i,j})\left\|\x_i-\x_j\right\|].
\end{array}
\end{equation}

Obviously, (\ref{eqn:minproblem}) is nonconvex, which cannot be solved easily. However, we can relax the problem to two convex optimization problems by using the SDP relaxation techniques as proposed in \cite{sdpipsn} and \cite{fsdp}, which are referred as FullSDP and ESDP, respectively.

Define
\begin{equation}
\begin{array}{l}
 X=[x_1,x_2,...,x_{n+m}]\in R^{2\times (n+m)}\\
 Y=X^TX.\\
 \end{array}
 \end{equation}
We also define:
\begin{equation}
\begin{array}{l}
\gamma_{i,j}=g_{i,j}^2\\
g_{i,j}=\left\|\x_i-\x_j\right\|.
\end{array}
\end{equation}
Then we can write (\ref{eqn:minproblem}) as follows:
\begin{equation}
\begin{array}{ll}\label{eqn: non-convex}
\mbox{min} &\sum_{i<j;(i,j)\in E} [\gamma_{i,j}-2(l_{i,j}+u_{i,j}) g_{i,j}]\\
\mbox{s.t.} & \gamma_{i,j}=Y_{ii}+Y_{jj}-2Y_{ij}\\
& \gamma_{i,j}=g_{i,j}^2\\
& Y=X^TX\succeq 0.\\
\end{array}
\end{equation}

By performing the SDP relaxation, we relax (\ref{eqn: non-convex}) to a convex program as follows:
\begin{equation}
\begin{array}{ll}\label{eqn:FULLSDP 1}
\mbox{min}_{\gamma, g, Z} &\sum_{i<j;(i,j)\in E} [\gamma_{i,j}-2(l_{i,j}+u_{i,j}) g_{i,j}]\\
\mbox{s.t.} & \gamma_{i,j}=Y_{ii}+Y_{jj}-2Y_{ij}\\
& \gamma_{i,j}\ge g_{i,j}^2\\
&
Z=\left(
  \begin{array}{cc}
    I_2 & X \\
    X^T & Y \\
  \end{array}
\right)\succeq 0\\
\end{array}
\end{equation}
where $I_2$ is the $2\times 2$ identity matrix.

It is worth noting that the anchor part of $Z$ is known, therefore it is also in the constraint. Hence, we formulate the localization program in (\ref{eqn:FULLSDP 1}).

When the problem is large, the SDP formulation might be slow \cite{cop}. Based on the work of \cite{fsdp}, we can further relax it into an ESDP formulation:
\begin{equation}
\begin{array}{ll}\label{eqn:ESDP 1}
\mbox{min}_{\gamma,g,Z} &\sum_{i<j;(i,j)\in E} [\gamma_{i,j}-2(l_{i,j}+u_{i,j}) g_{i,j}]\\
\mbox{s.t.} & \gamma_{i,j}=Y_{ii}+Y_{jj}-2Y_{ij}\\
& \gamma_{i,j}\ge g_{i,j}^2\\
&
Z=\left(
  \begin{array}{cc}
    I_2 & X \\
    X^T & Y \\
  \end{array}
\right)\\
& Z_{(1,2,i,j)}\succeq 0 \mbox{    } \forall (i,j)\in E\\
\end{array}
\end{equation}
where $Z_{(1,2,i,j)}$ denotes the principal submatrix of $j$ consisted of row and column $1,2,i,j$.

 Both (\ref{eqn:FULLSDP 1}) and (\ref{eqn:ESDP 1}) can be solved by standard SDP solvers such as SeDuMi or SDPT3 in a centralized way \cite{ksdp}. We choose YALMIP \cite{YALMIP} as the programming interface.

\begin{remark}
In practice, one might want to add different weights to different terms in objective according to the confidence he has on each measurement. For instance, we can give a lower weight to NLOS part if we have prior statistics on NLOS measurements.
\end{remark}

In the next section, we are going to discuss one extension of above model, i.e., the situation in which there are uncertainties in the anchor positions. We show that a similar SDP model can be formulated to solve this problem.

\section{Localization in NLOS Environment considering Anchor Errors}

In the model where uncertainties exist for anchor positions, we assume that the true anchor positions are within a certain ball around the estimated ones, namely, for each anchor $j$
 \[\|\x_j-\bar \x_j\| \le {u_j}\]
where $\bar \x_j $ is the estimated value while $\x_j$ is the true one.
In our cases, $u_j$ is given, and it usually comes from the confidence in the measurement uses.

In this case, in addition to the objective in (\ref{eqn:minproblem}), we also add the anchor error in the objective. And the anchor positions become variables as well.

We formulate the problem with anchor position uncertainty as follows:
\begin{equation}
\begin{array}{ll}\label{eqn:minproblem2}
\mbox{min}{x} & \sum_{i<j:(i,j)\in E}[\left\|\x_i-\x_j\right\|^2-2(l_{i,j}+u_{i,j})\left\|\x_i-\x_j\right\|]\\
& + \sum_{j=m}^{m+n}\left\|\x_j-\bar \x_j\right\|^2\\
\end{array}
\end{equation}
where the second term is about the anchor uncertainty. Note that one can also add some weight to each term, denoting the different confidence level one has for each measurement.

By using the same technique, we can relax (\ref{eqn:minproblem2}) into a convex program. Again, define
\begin{equation}
\begin{array}{l}
 X=[x_1,x_2,...,x_{n+m}]\in R^{2\times (n+m)}\\
 Z=\left(
     \begin{array}{cc}
       I_2 & X \\
       X^T & X^TX \\
     \end{array}
   \right)
 \end{array}
 \end{equation}
and
\begin{equation}
\begin{array}{l}
\gamma_{i,j}=g_{i,j}^2\\
g_{i,j}=\left\|\x_i-\x_j\right\|.
\end{array}
\end{equation}

Then we can write (\ref{eqn:minproblem2}) as follows:
\begin{equation}
\begin{array}{ll}\label{eqn: non-convex2}
\mbox{min}_{\gamma,g,Z} &\sum_{i<j;(i,j)\in E} [\gamma_{i,j}-2(l_{i,j}+u_{i,j}) g_{i,j}]\\
&\mbox{  }+\sum_{j=n+1}^{n+m}[Z_{j+2,j+2}-2Z_{1:2,j+2}^T\bar\x_j]\\
\mbox{s.t.} & \gamma_{i,j}=Z_{i+2,i+2}+Z_{j+2,j+2}-2Z_{i+2,j+2}\\
& \gamma_{i,j}=g_{i,j}^2\\
& Z=\left(
     \begin{array}{cc}
       I_2 & X \\
       X^T & X^TX \\
     \end{array}
   \right)\succeq 0.\\
\end{array}
\end{equation}

By performing the SDP relaxation, we relax (\ref{eqn: non-convex2}) to a convex program as follows:
\begin{equation}
\begin{array}{ll}\label{eqn:FULLSDP 2}
\mbox{min} &\sum_{i<j;(i,j)\in E} [\gamma_{i,j}-2(l_{i,j}+u_{i,j}) g_{i,j}]\\
&\mbox{   }+\sum_{j=n+1}^{n+m}[Z_{j+2,j+2}-2Z_{1:2,j+2}^T\bar\x_j]\\
\mbox{s.t.} & \gamma_{i,j}=Z_{i+2,i+2}+Z_{j+2,j+2}-2Z_{i+2,j+2}\\
& \gamma_{i,j}\ge g_{i,j}^2\\
& Z_{(1,2)}=I_2\\
& Z\succeq 0.\\
\end{array}
\end{equation}

By using the same method as in (\ref{eqn:ESDP 1}), we also get the ESDP relaxation to this case:
\begin{equation}
\begin{array}{ll}\label{eqn:FULLSDP 2}
\mbox{min} &\sum_{i<j;(i,j)\in E} [\gamma_{i,j}-2(l_{i,j}+u_{i,j}) g_{i,j}]\\
&+\sum_{j=n+1}^{n+m}[Z_{j+2,j+2}-2Z_{1:2,j+2}^T\bar\x_j]\\
\mbox{s.t.} & \gamma_{i,j}=Z_{i+2,i+2}+Z_{j+2,j+2}-2Z_{i+2,j+2}\\
& \gamma_{i,j}\ge g_{i,j}^2\\
& Z_{(1,2)}=I_2\\
& Z_{(1,2,i,j)}\succeq 0\mbox{    }\forall (i,j)\in E.\\
\end{array}
\end{equation}

\section{Numerical Results}
\begin{figure}[t]
\centering
\includegraphics[width=3.5in]{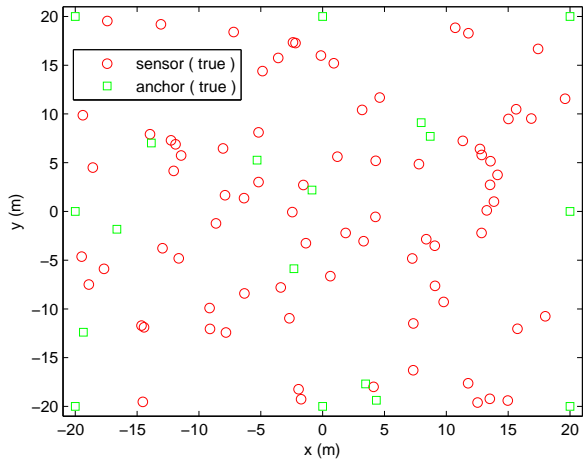}
\caption{The estimated sensor positions} \label{fig:anchor_sensor_place}
\end{figure}

In this section, simulation results are presented and analyzed.
The performance evaluation focuses on the position estimation accuracy of the proposed algorithm.
We consider a 2-dimensional region with a size of 40 m $\times$ 40 m.
There are totally 18 anchors locating in the area.
Eight of them are located at the boundary (20, 20)m, (-20,20)m, (20,-20)m, (-20,-20)m, (0,0)m, (-20,0)m and (0,-20)m, while the remaining ten anchors are randomly deployed in the area.
In this simulation they are localized at (4.3416,-19.3696)m, (-19.3458,-12.3970)m, (3.4767,-17.6967)m, (-5.2972,5.2580)m, (8.7053,7.7067)m, (-16.6368,-1.8257)m, (-2.3268,-5.8699)m, (-13.8557,7.0257)m, (7.9685,9.1003)m and (-0.8646,2.1936)m.
Then, we deploy 80 sensors in the field, but the number of sensor is too large to be listed, so we omit the listing of their coordinates.
Nevertheless, the configuration is shown in Fig. \ref{fig:anchor_sensor_place}.
For the sake of simplicity, we assume that all the sensors and anchors can see each others, i.e., the fully connected situation is considered.

\begin{figure}[t]
\centering
\includegraphics[width=3.5in]{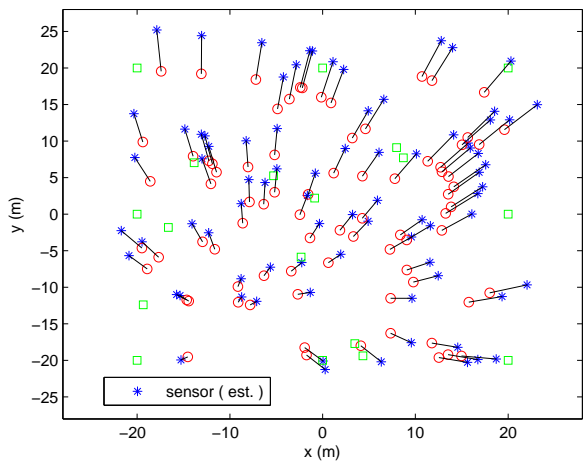}
\caption{The configuration of the network} \label{fig:est_place}
\end{figure}

We follow the noise model of (\ref{eqn:noise_model}) with $n_i$ a normally distributed variable with noise power -40dB and ${\delta _i}$ being a uniformly distributed random variable drawn from $[0,0.5]$.
That means all measurements contains an NLOS error.
Our proposed SDR formulation of (\ref{eqn:FULLSDP 1}) is applied to find the estimated sensor positions and the result is shown in Fig. \ref{fig:est_place}.
The average mean square position error is 6.6613m$^2$.
From the figure we see that the proposed method can provide good estimation by mitigating the effects of NLOS measurements.
The solution can act as an initial guess for other numerical search to obtain better estimate.

\section{Conclusions}
A semi-definite programming based node localization algorithm in NLOS environments
for UWB wireless sensor networks has been proposed in this paper.
The problem of node localization in the presence of anchor position uncertainty has been
approximated by a convex optimization problem using the SDP relaxation technique.
Given a mixture of LOS and NLOS range measurements, our method is
applicable in both cases without discarding any range information.
Simulation results demonstrate the effectiveness of our method.

\section*{Acknowledgement}
We would like to thank Dr. W.-K. Ma from the Chinese University of Hong Kong for his
valuable suggestions concerning this work.


\begin{thebibliography}{1}

\bibitem{apit}
T. He, C. Huang, B.M. Blum, J.A. Stankovic, and T. Abdelzaher, "Range-free localization schemes for large scale sensor networks," in\emph{ Proc. ACM MobiCom}, San Diego, CA, Sept. 2003, pp.81-95.
\bibitem{mdtoa}
J. Luo, H. V. Shukla, and J.-P. Hubaux, "Non-interactive location surveying for sensor neworks with mobility-differentiated ToA," in\emph{ Proc. IEEE INFOCOM}, Barcelona, Spain,
Apr. 2006, pp.1-12.
\bibitem{chanNLOS}
Y.T. Chan, W.Y. Tsui, H.C. So, and P.C. Ching, "Time-of-arrival based
localization under NLOS conditions," \emph{IEEE Trans. Veh. Technol.,} vol. 55,
no. 1, pp. 17-24, Jan. 2006.
\bibitem{congNLOS}
L. Cong and W. Zhuang, "Nonline-of-sight error mitigation in mobile
location," \emph{IEEE Trans. Wireless Commun.,} vol. 4, no. 2, pp. 560¨C573,
Mar. 2005.
\bibitem{wangNLOS}
W. Wang, Z. Wang, and B. O'Dea, "A TOA-based location algorithm
reducing the errors due to non-line-of-sight (NLOS) propagation," \emph{IEEE
Trans. Veh. Technol.,} vol. 52, no. 1, pp. 112-116, Jan. 2003.
\bibitem{caffery}
S. Venkatraman, J. Caffery, and H.R. You, " A novel ToA location algorithm using LoS range
estimation for NLoS environments," \emph{ IEEE Trans. Veh. Technol.,} vol. 53, no. 5, pp.1515-1524.


\bibitem{sdpipsn}
P. Biswas and Y. Ye, "Semidefinite programming for ad hoc wireless
sensor network localization," in \emph{Proc. ACM IPSN}, pp. 46-54, 2004.
\bibitem{fsdp}
Z. Wang, S. Zheng, Y. Ye, and S. Boyd, "Further relaxations of
the semidefinite programming approach to sensor network localization,"
\emph{SIAM J. Optim.,} vol. 19, no. 2, pp. 655--673, Jul. 2008.
\bibitem{cop}
S. Boyd and L. Vandenberghe, Convex Optimization. Cambridge University
Press, 2004.
\bibitem{ksdp}
K.W.K. Lui, W.-K. Ma, H.C. So, and F.K.W. Chan, "Semidefinite
programming algorithms for sensor network node localization
with uncertainties in anchor positions and/or propagation speed," \emph{ IEEE
Trans. Signal Process.,} vol. 57, no.2, pp.752-763, Feb. 2009.
\bibitem{YALMIP} J. L\"{o}fberg, "YALMIP : a toolbox for modeling and optimization in MATLAB," {\sl Proc. Int. Symp. CACSD}, pp. 284--289, Taipei, Taiwan, Sep. 2004.

\end{thebibliography}
\end{document}